# Monte Carlo Simulations of Infection Spread in Indoor Environment

*Rahul Sheshanarayana and Prateek K. Jha\**
*Department of Chemical Engineering, IIT Roorkee, Roorkee, Uttarakhand, 247667, India*
*E-mail: prateek.jha@ch.iitr.ac.in*

## Abstract

The dynamics of infection spread in populations has received popular attention since the outbreak of Covid-19 and many statistical models have been developed. One of the interesting areas of research is short-time dynamics in confined, indoor environments. We have modeled this using a simple Monte Carlo scheme. Our model is generally applicable for the peer-to-peer transmission case, when the infection spread occurs only between an infected subject and a healthy subject with a certain probability, i.e., airborne and surface transmission is neglected. The probability of infection spread is incorporated using a simple exponential decay with distance between the subjects. Simulations are performed for the cases of (1) constant subject population and (2) variable subject population due to inflow/outflow. We specifically focus on the large fluctuations in the dynamics due to finite number of subjects. Results of our study may be useful to determine social-distancing guidelines in indoor contexts.

## Introduction

Growth kinetics of infectious diseases have been of great interest to scientists and mankind since very long, but it has gained tremendous attention after the Covid-19 outbreak. A large number of models focusing on the prediction of Covid-19 infections in different countries have been developed, which have mixed success due to an enormous number of factors affecting the disease spread, a lack of understanding of infection mechanism, and inadequate testing rates in many countries. However, social-distancing and lockdown strategies have proved successful in the containment of the spread. A global vaccination drive is currently underway that is expected to alleviate the situation. However, many questions still remain for the future course of action for this and other infection spread. One of the major questions pertain to the mitigation of infection spread in indoor environments with high footfalls of "subjects" (people), e.g., classrooms, public transport, etc. Three lines of thought have emerged in this area: (1) minimize interaction between subjects by promoting "work at home" wherever possible, online education and shopping, etc., (2) improve airflow and ventilation in indoor settings, (3) compulsory use of protective masks and social-distancing measures in indoor settings. It goes without saying that such measures are far from achievable in many settings, especially in classrooms and public transport.

From a mathematical perspective, infection spread in an indoor setting is a stochastic process affected by the subject movement and interaction between subjects, vulnerability of subjects, composition of the indoor environment, and the nature and extent of the infection. Few observations are in order. Since all these factors are highly variable, it is probably not worthwhile to predict the rate of infection. Instead, one may judge the best-case and worst-case situations and the effectiveness of a mitigation strategy by comparing different scenarios. Both the average rate of infection and its fluctuations around the average are important for policy decisions. Such fluctuations are expected to be large in most indoor settings with small, finite number of subjects. Finally, unless airborne or surface transmission is the sole mechanism of spread, increasing distance between subjects is bound to result in lower infection rates. In fact, this explains the fast spread of infection within members of a family, despite the use of preventive measures. Based on these observations, we have attempted to develop a general stochastic model of infection spread in confined settings that only considers peer-to-peer transmission between subjects with the probability of infection decaying with distance between subjects. Monte Carlo simulations performed on this model provides useful insights on infection spread in indoor settings. It is important to point out that the use of Monte Carlo simulations to model infection spread is not entirely new and several models exist [1], [2], [3]. Though the research conducted in this area address the problem to a great extent, they are either quantitative/analytical in nature, or applied to a certain infection in particular. The proposed model suggests a stochastic process that ensures compatibility with variation in human tendencies, which, in turn, affects the nature of infection spread. Also, it is based on robust parameters, hence motivating one to study different circumstances occurring due to the same infection, or cases involving multiple infections independently.
.
**Methodology**

Figure 1 shows the schematic of our model. At time $t = 0$, $n_0$ subjects are assumed to be present in the room at random locations, a fraction $f_0$ of which are infected. The initial population density of the room is defined $\rho_0 = n_0/A$ where $A$ is the area of the room. In a timestep $\tau_s$, the subjects move in a random direction defined as the random choice of angle

$\theta \in [0, 2\pi]$ with a fixed step size value of 1 ft, equivalent of one typical stride. The subjects formed $(n^2 - n)/2$ pairs, which were grouped into a 'trivial' and a 'non-trivial' group after every timestep. The trivial group consists of subjects who are both not infected or both safe. The non-trivial group has an infected and a safe subject. The probability of the safe subject of the non-trivial group to get infected in a particular timestep is assumed to be

$$p = p_0 \exp(-\kappa d^2) \tag{1}$$

where $d$ is the Euclidian distance between the subjects. In our numerical implementation, random numbers $\alpha \in [0,1]$ is generated for the non-trivial pairs and the infection occurs when $p < \alpha$. Here, coefficient $\kappa$ signify the decay rate and $p_0$ is the maximum probability obtained on physical contact between subjects ($d = 0$). In our numerical implementation, a random number Though eq. (1) captures the basic feature of probability decay with distance, one may argue the use of other functions to signify the decay of infections that will of course depend on the type of infection. The subject infected in a particular timestep may require an incubation time $\tau_i$ to be able to infect others.

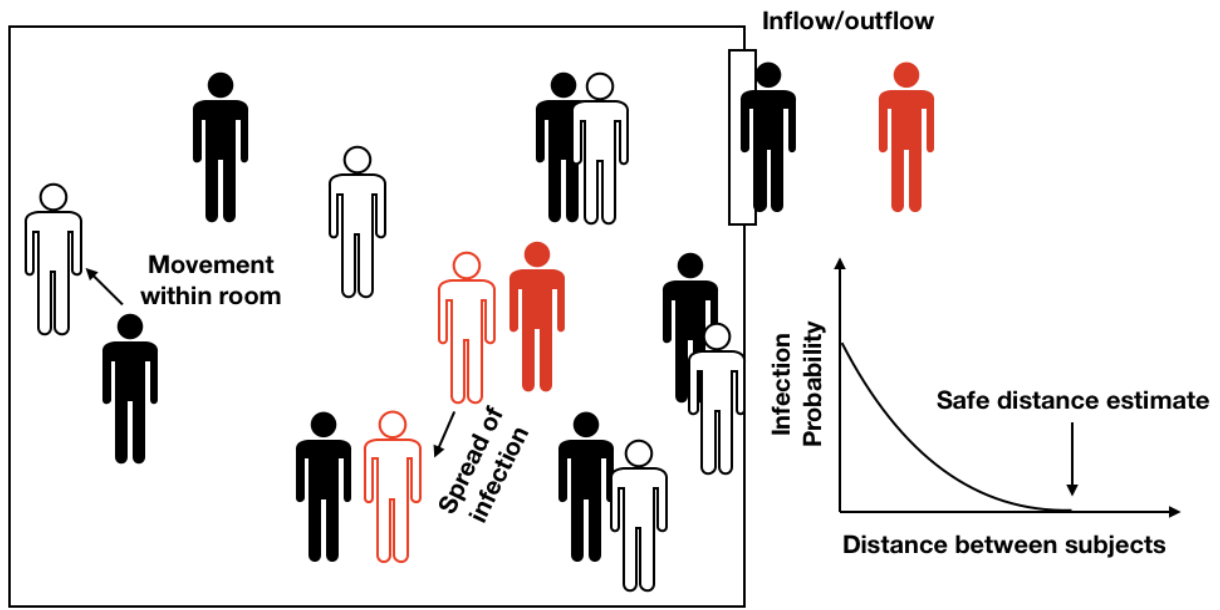

*Figure 1: Schematic of the model. Subjects are confined in a rectangular room with a door (top right) for possible inflow/outflow. Filled and open person representation indicate the old and new positions (after some time) of subjects. Red color indicates infected subjects. Typical plot of infection probability of subjects with distance are also shown.*

**Results and Discussion**

Simulations are performed for several scenarios to observe the pattern of infection spread. However, due to a lack of reliable data on infection spread in indoor settings, no effort is made to fit the model to available results. It should therefore be noted that the results of our study are applicable for our choice of model parameters and the infection probability (eq. 1) and not generally applicable. However, we have attempted to use realistic values of the model parameters. Simulations are always performed for 100 different seed values and the average and standard deviations of the infection profile is plotted. The choice of only 100 different seed values is made with a purpose. The context in which the models may be applied are not universal and therefore it is incorrect to imagine that the model parameters of our simulations can be applicable and tested for a large number of similar situations, e.g., room dimensions and the average number of subjects. At best, this model may be applied within particular institutes or enterprises or public transport systems with many similar rooms and footfalls. The statistical-mechanical framework of the Monte Carlo simulations adopted in this work is therefore not quite rigorous and are not considered in the thermodynamic limit, $n \to \infty$ and large room dimensions.

We first evaluate the 'social distance' criteria prescribed by the World Health Organization (WHO) for a closed environment in the context of Covid-19, which is a minimum safe distance of $d_{safe} = 6$ ft [4]. Assuming that the airborne and surface transmission is not significant, one may assume that the infection probability is very small for $d = d_{safe}$ in eq. (1). Further, we may imagine a high infection probability at low distance, e.g., $d = 1$ ft. Assuming $p(1 \text{ ft}) = 0.2$ and $p(d_{safe}) = 10^{-7}$ in eq. (1), we can estimate $k = 0.4145$ ft$^{-2}$ and $p_0 = 0.3027$. 3

Figure 2 shows the results on a **public transport case** assuming no inflow/outflow. Since the average movement of individuals are small except during onboarding and deboarding, we may imagine $\tau_s$ to be large ($\approx 5$ min), in which case 26% average infection would occur after 50 minutes starting with 5% infected subjects for the case of 1 subject in 50 sq. ft., $\rho_0 = 1/50$ ft$^{-2}$, provided that we use $k, p_0$ values estimated above and the incubation time is negligible ($\tau_i = 0$). The standard deviations of the fraction of infected individuals are quite high due to small system size. In this case, the best case and worst-case scenario after 60 min are 12% and 40%. It is worth pointing out that the rate of increase actually decreases with time even while the number of infected subjects increases. This is because of small movements of subjects in a time step that results in fast infection of subjects in their small neighborhood but they have a much slower approach towards distant safe subjects.

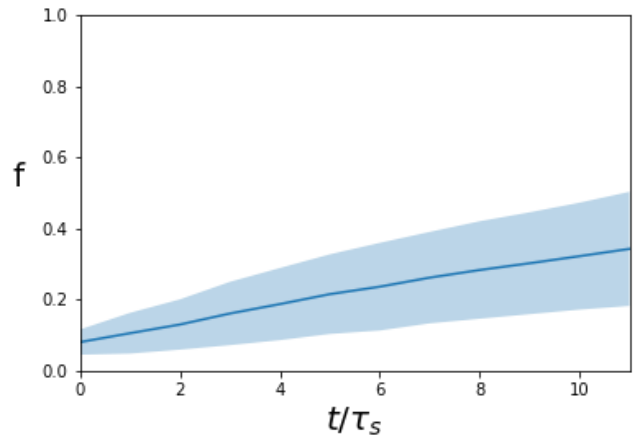

*Figure 2: Fraction of infected population against time for the **public transport case**. The model parameters are $k = 0.4145\ ft^{-2}$, $p_0 = 0.3027$, $n_0 = 20$, $f_0 = 0.05$, $\rho_0 = 1/50\ ft^{-2}$, $\tau_i = 0$, $\tau_s = 5$ no inflow/outflow. The bold line indicates the mean value and the shaded region indicates the range of possible profiles.*

Figure 3 below shows the results for a **classroom case** at the beginning when the class of students are constantly entering into a classroom at a predefined rate with $\Delta n$ students entering in time $\tau_s$. Starting with 4% infected subjects, an average 75% subjects get infected in 30 timesteps for $\rho_0 = 1/5\ ft^{-2}$ and $\Delta n = 3$, provided that we use $k, p_0$ values estimated above and the incubation time is negligible ($\tau_i = 0$). Again, the standard deviations of the fraction of infected individuals are quite high due to small system size. In this case, the best case and worst-case scenario after 30 timesteps are 40% and 100%. Unlike the public transport case, the rate of infection increases with time, since the population density is rather high and there is inflow of subjects in the room. Still, the actual magnitude of infection spread in a classroom setting would be smaller than what is apparent from Figure 3. This is because, for a typical value of $\tau_s \approx 1$ min, $\Delta n$ will not remain constant after few timesteps and most students would have entered the classroom in $\sim 5 - 10$ min (timesteps).

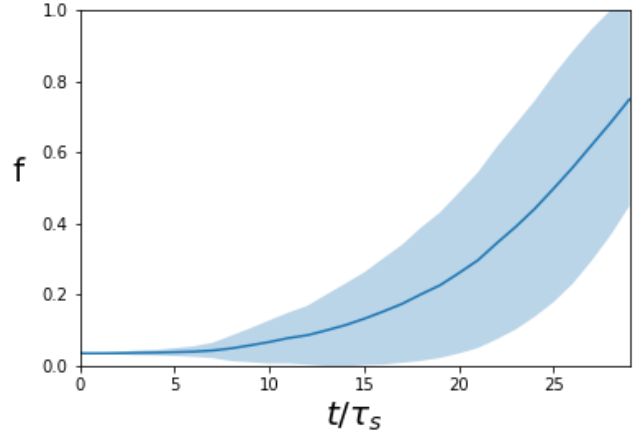

*Figure 3: Fraction of infected population against time for the **classroom** case. The model parameters are $k = 0.4145\ ft^{-2}$, $p_0 = 0.3027$, $n_0 = 30$, $f_0 = 0.04$, $\rho_0 = 1/5\ ft^{-2}$, $\tau_i = 0$, constant inflow of students with $\Delta n = 3$. The bold line indicates the mean value and the shaded region indicates the range of possible profiles.*

As apparent from the above two cases, the population density and room dimensions are important factor that determines the infection growth. We have therefore performed a systematic study of their effects in Figure 4. Interestingly, the cases of infection rose at a harmful rate in situations where the number of subjects considered initially ($n_0 = \rho_0 A$) was higher (Figures 4b,4d,4f), given the same $\rho_0$ and $A$. The worst scenario occurred in the case of situation considered in Figure 4 (b), i.e., when all the subjects got infected within 5 timesteps with a high certainty.

The slope of the curves below are decreasing in nature, implying that after a short period of time, the number of subjects getting infected shows least amount of change with the timesteps. Hence accommodating subjects in such a setting for a time until the value of $f$ is under a threshold is a viable decision to make. This threshold for $f$, and consequently the maximum time for a gathering can be inferred by a plot, such as the below, generated using confident parameter values known for the situation in hand. Also, the certainty of $f \to 1$ towards the end of the experiments with the same number of timesteps starts to decrease as $\rho_0$ is decreased. Similar trend is shown by the subsequent two situations as well.

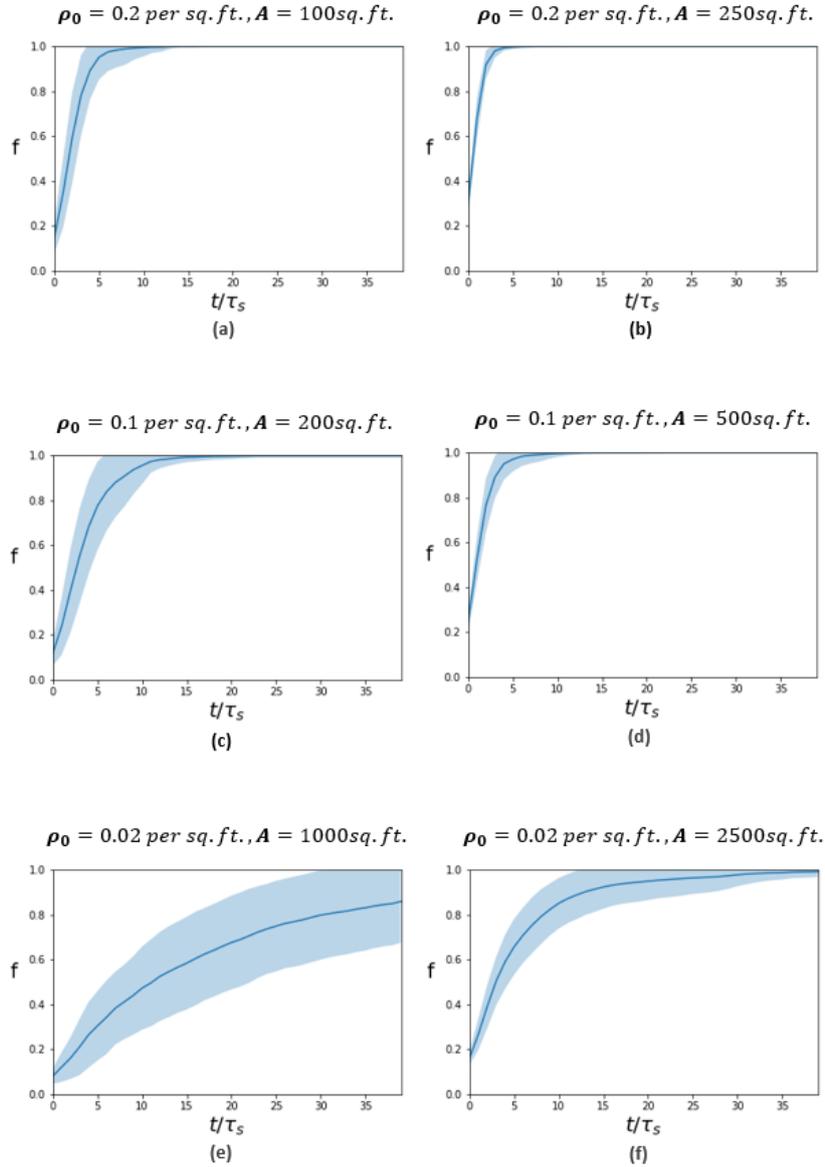

*Figure 4: Effect of population density. (a),(b) are for **high population density**, (c), (d) are for **intermediate population density**, and (e), (f) are for **low population density**. (b), (d), (f) have same population density as (a), (c), (f), respectively, but have larger room area. Other model parameters for all plots are $k = 1/9 \, ft^{-2}$, $p_0 = 0.08$, $f_0 = 0.2$, and $\tau_i = 0$. No influx of subjects is assumed.*

*A varying number of observations per timestep*

In this experiment (Figure 5), the number of observations during the experiment was varied with a constant rate of influx. The value of the influx rate ($\alpha$) was diversified throughout experiments conducted under low, intermediate, and high-density cases. A parameter $\Delta n$ (= $\alpha \Delta t$) was set to ensure a constant inflow of observations after each timestep.

In this case, the slope of the curves gradually increases until a value of $f$ is reached after which the slope remains somewhat constant at a relatively high value and then starts to decrease. One may infer that, given such a trend, accommodating subjects in such a scenario can be possible only for a short period of time initially - until the slope increases very gradually.

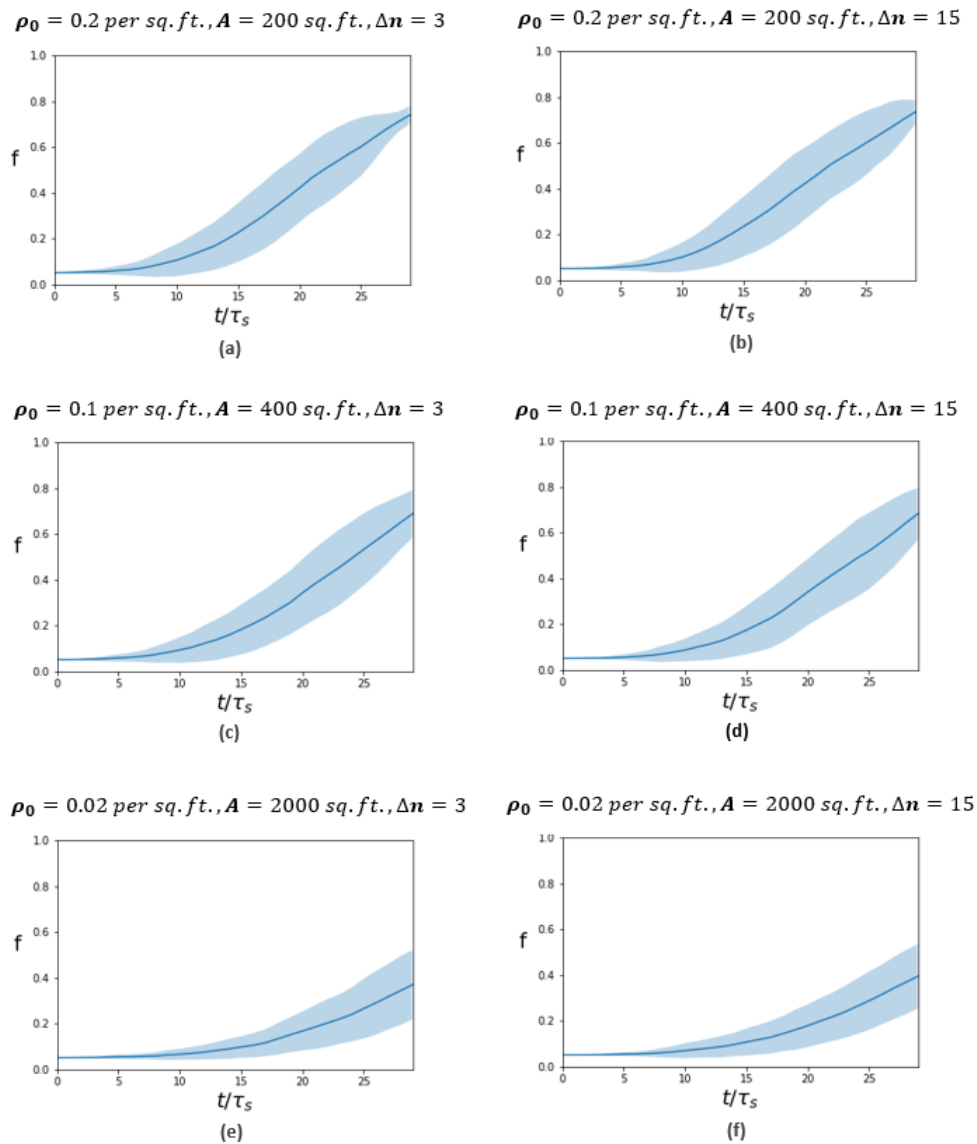

Figure 5: The parameter values used in common to generate all the above plots are $k = 1/9\ ft^{-2}$, $p_0 = 0.2$, $f_0 = 0.08$, $\tau_i = 0$, $\tau_s = 1$

### The effect of parameter $\tau_i$

The constant $\tau_i$ (incubation time) here is the parameter that was incorporated to control the time up to which a subsect could remain harmless before starting to spread the infection it acquired. After getting infected, a subject doesn't begin to spread the disease until a certain period has elapsed in real life. This time depends on many factors such as immunity of the subject, viral load on the subject, etc. The study of the effect of $\tau_i$ was again carried out under three different scenarios - high, intermediate, and low population density. Unlike in the previous two cases, the experiments were analyzed using plots between $f$ and active time ($t^*$), as shown in Figure 6. The proposed way of analysis ensured that the infection-growth is monitored at non-trivial times, i.e., neglecting times when the infected observations are going through their incubation period - a time when they are not spreading the disease prevalent in/acquired by them. To clarify, all the plots below are generated for the same duration of 40 timesteps.

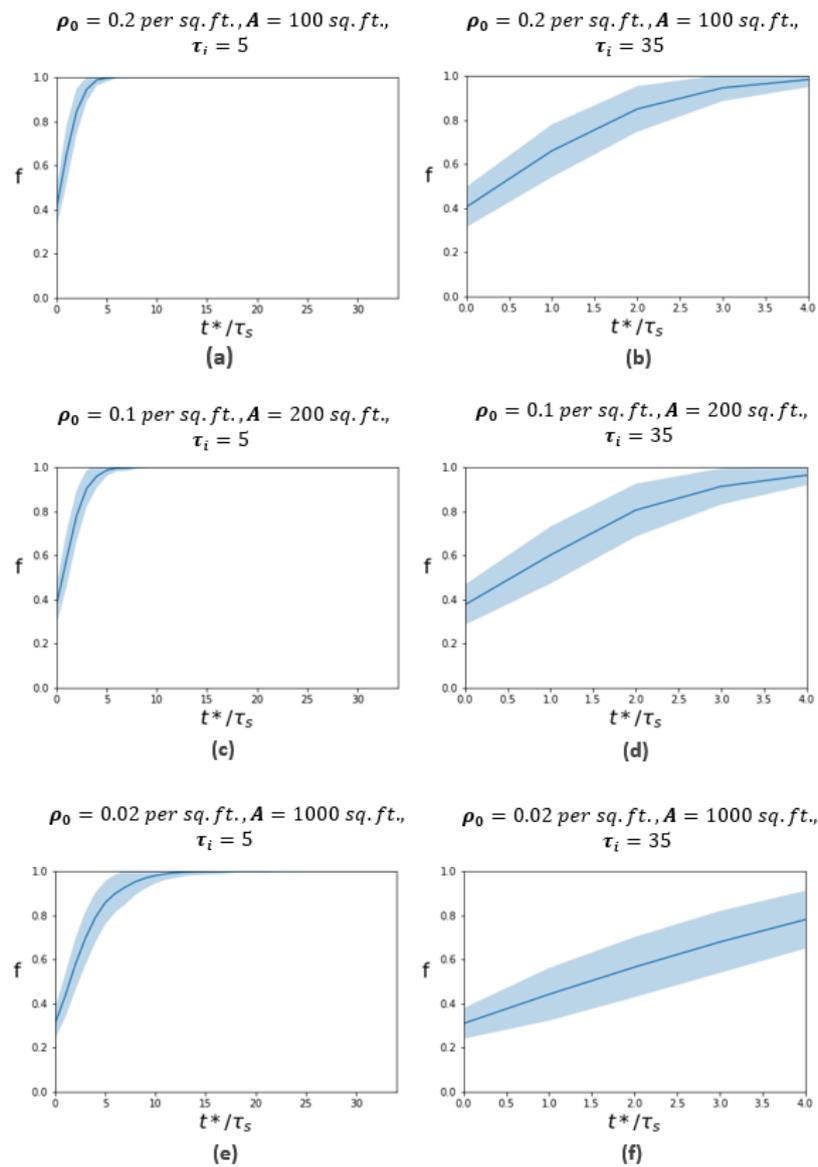

Figure 6: The parameter values used in common to generate all the above plots are $k = 1/9\ ft^{-2}$, $p_0 = 0.08$, $f_0 = 0.2$, $\tau_i = 0$, $\tau_s = 1$

**Conclusion**

A peer-to-peer transmission model has been developed for the spread of infections in confined spaces. Monte Carlo simulations of the model provides interesting insights that may potentially be useful in developing mitigation strategies for containing such infections. Specifically, two scenarios ('public transportation case' and 'classroom case') has been studied in detail, which provides a good understanding of infection spread in these cases.